\documentclass[11pt]{article}
\usepackage{moriond,epsfig}

\def\mpp{M_{p\bar{p}}}
\def\mb{{M_{bc}}}
\def\de{{\Delta{E}}}
\def\bp{{B^{+}}}
\def\bz{{B^{0}}}
\def\pp{{p\bar{p}}}
\def\ppk{{p\bar{p}K^+}}
\def\pppi{{p\bar{p}\pi^+}}

\def\ppks{{p\bar{p}K_S^0}}
\def\ppkst{{p\bar{p}K^{\star +}}}
\def\ppksz{{p\bar{p}K^{\star0}}}
\def\plpi{{p\bar{\Lambda}\pi^-}}
\def\plk{{p\bar{\Lambda}K^-}}
\def\psigpi{{p\bar{\Sigma}^0\pi^-}}

\def\alamc{\bar\Lambda_c}

\begin{document}
\vspace*{4cm}
\title{Measurement  of Baryonic $B$ Decays}
\author{ M. Z. Wang }

\address{ {\rm on behalf of the Belle Collaboration}\\
Department of Physics, National Taiwan University, Taipei, Taiwan, 
R.O.C.}

\maketitle\abstracts{
Recent results of baryonic $B$ decays from Belle are reported.  
This study is done by a  78 fb$^{-1}$  data sample,
consisting of $85.0 \pm 0.5$ million $B\overline{B}$ pairs,
collected by the Belle detector 
at the KEKB asymmetric energy $e^+e^-$ (3.5 on 8~GeV) collider. 
The results reported here include the first observation of the two-body
decay $\bz \to p\alamc$, the first hyperonic decay $\bz \to \plpi$,  and first
observations of $\bp \to \pppi$, $\bz \to \ppks$, and $\bp \to \ppkst$.
}
\section{Introduction}

The Belle collaboration recently reported 
the observation of $B^+ \to p\bar{p}K^+$\footnote{Throughout this report, 
inclusion of charge conjugate mode is always implied
unless otherwise stated.}~\cite{ppk}, 
which is the first known example of $B$ meson
decay to charmless final states containing baryons. 
The three-body decay rate is larger than the rate for two-body decays
(such as $B\to p\bar p$~\cite{2body}),
We continue this study with a larger data set and
search for other related baryonic decay modes.  In the Standard Model, 
these decays proceed via $b\to u (c)$ tree and $b\to s (d)$ 
penguin diagrams. They may be used to 
search for direct $CP$ violation and 
test our theoretical understanding of 
rare decay processes involving baryons~\cite{Theory}.

Belle~\cite{Belle} is a general purpose detector
operating at the KEKB asymmetric $e^+e^-$ collider.   
Data sample used here  consists of $85.0 \pm 0.5$ 
million $B\overline{B}$ pairs or $\sim$78 fb$^{-1}$ data set 
collected on the $\Upsilon(4S)$ resonance.    
   
\section{Analysis Procedure}
The biggest challenge of observing rare $B$ decay processes 
is to fish out a few 
signal events from a huge sample of background events. For example, 
after the trigger and hadronic 
event pre-selection, there are 
about 170 million $e^+e^- \to q\bar{q}$ continuum events
and 85 million $B\bar{B}$ events left for the 78 fb$^{-1}$ 
data set. It is a tough job to reject all of the background. The following
is a brief description of the procedure of a typical analysis.

\subsection{Signal identification}
Since the center-of-mass energy is set to match 
the $\Upsilon(4S)$ resonance and  $\Upsilon(4S)$ decays into a  
$B\bar{B}$ pair, 
we can use the following two kinematic variables to 
identify the reconstructed
$B$ meson candidates: the beam-energy constraint mass, $M_{bc} =
\sqrt{E^2_{beam}-p^2_B}$, and the energy difference, $\Delta{E} =
E_B - E_{beam}$, where $E_{beam}$, $p_B$ and $E_B$ are the beam energy, the
momentum and energy of the reconstructed $B$ meson in the rest frame of 
$\Upsilon(4S)$, respectively. 

The resolution of $M_{bc}$ is about 3 MeV/$c^2$ which is due to the spread 
of the beam energy. Typically, the resolution of  $\Delta{E}$ is 
about 10 MeV
for final states with charged particles only. 

\subsection{Background suppression}
The generic $B$ decay is mainly via the $b \to c$ transition which
normally has  more final state particles than those of the rare decay modes 
reported here, 
thus the background from generic B decays is much less than
that from the continuum process. Similar $B$ decay processes close to the
target mode should be checked carefully one by one because they might 
feed into the signal region.

For the continuum events, they have
quite different event topology (more back-to-back or jet-like) than that
of $B\bar{B}$ events (more spherical) in the $\Upsilon(4S)$ frame. 
We can select some shape variables  to form 
a Fisher discriminant in order to reject the  
continuum background.

The probability density functions (PDFs) of the Fisher discriminant and 
other uncorrelated kinematic variables 
(e.g. the angle between the $B$ flight direction 
and the beam direction),
are combined to form signal (background)
likelihood ${\cal L}_{\rm S (BG)}$, and a cut is then applied on
the likelihood ratio ${\cal LR} = {\cal L}_{\rm S}/({\cal L}_{\rm
S}+{\cal L}_{\rm BG})$ in order to fish out the signal.
%

\subsection{Yield determination}
After the optimization of selection cuts, events in the
candidate $\mb-\de$ region are used for yield determination.
This can be done either by un-binned likelihood fit
or binned fit. The signal PDFs
are normally a Gaussian function for $\mb$ 
and a double Gaussian for $\de$ with parameters determined by 
MC simulation. The background PDFs are Argus function for $\mb$
and a straight line for $\de$ with parameters determined by
sideband events or by continuum MC simulation.
If there is no evidence of signal events, one can use the fit results to 
estimate the expected background, 
and compare this  with the observed number of events  
in the signal region 
in order to set the upper limit on the 
yield at the 90\% confidence level~\cite{Conrad}.

\subsection{Systematic check}
If one is lucky to observe a new decay mode, the comparison between
signal MC simulation and data is necessary. However, most of 
the systematic 
studies are limited by the small statistics of the signal events. 
Therefore, control samples with large statistics and 
relevant to the study 
are checked in order to determine the systematic errors.
For example,
the systematic error due to the efficiency error of the
proton identification requirements
is studied with the  $\Lambda \to p \pi^-$ sample; 
the kaon identification is studied with the $D^{*+} \to D^0\pi^+$, 
 $D^0 \to K^-\pi^+$ sample; 
the tracking efficiency is studied with the
$\eta \to \gamma\gamma$ and $\eta \to \pi^+\pi^-\pi^0$ sample, etc..

\section{Recent results}
\subsection{Results of $\bz \to p\alamc$}
\begin{figure}[h]
\centering
\mbox{\psfig{figure=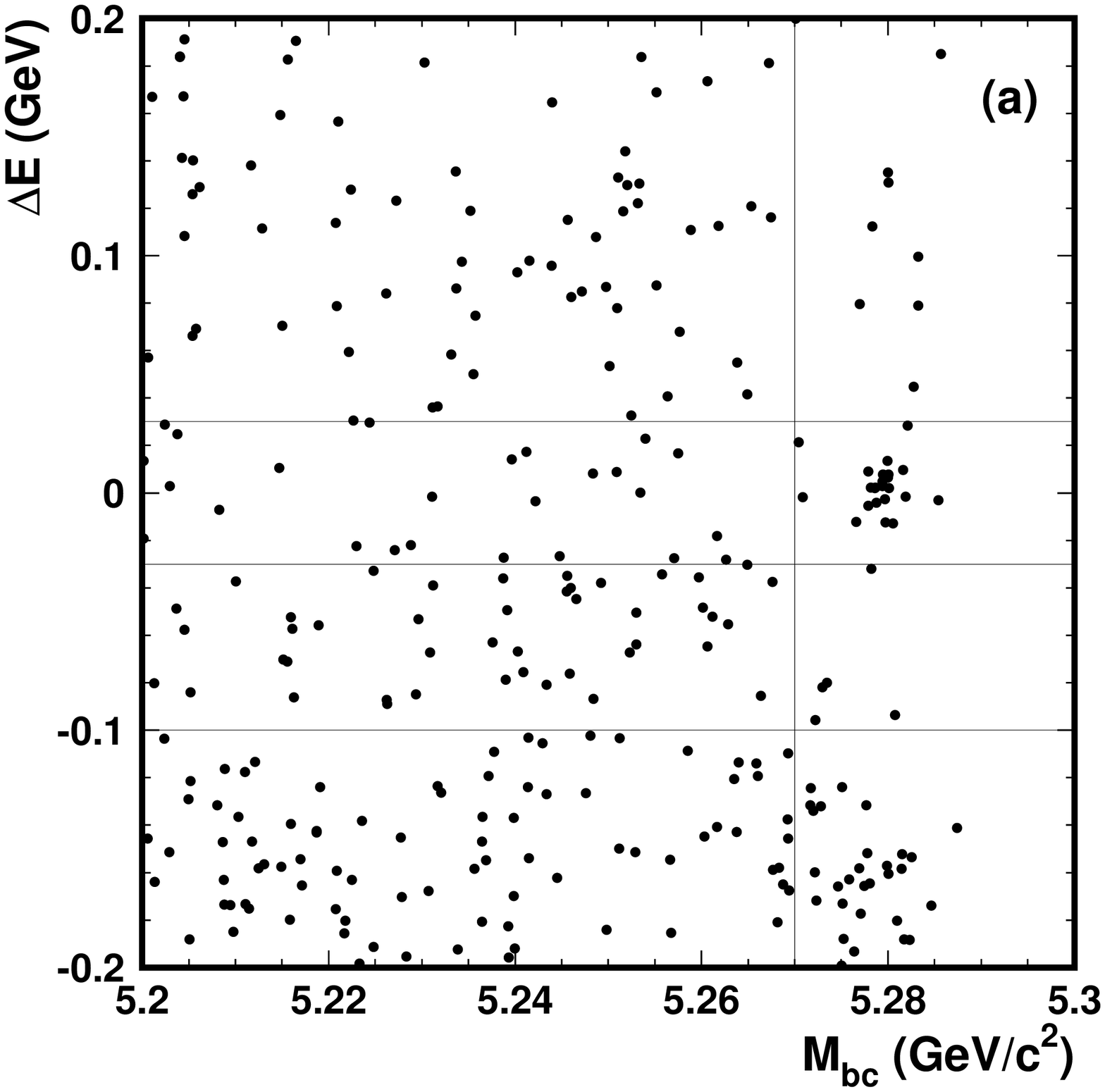,width=2.in}}
\mbox{\psfig{figure=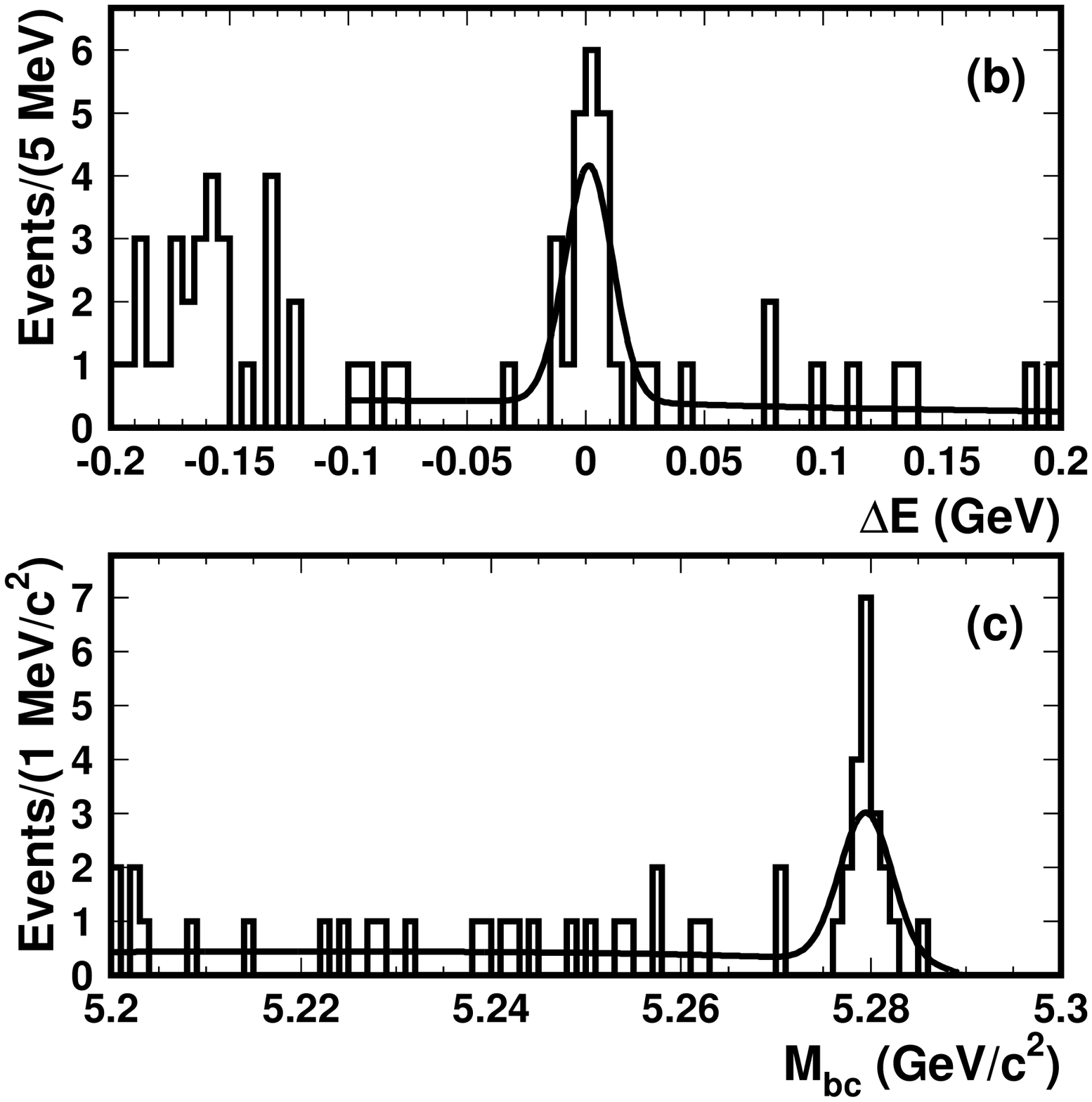,width=2.in}}
\centering
\caption{Candidate $\bz \to p\alamc$ events: (a) scatter plot of $\de$ 
versus $\mb$, (b) $\de$ distribution for $\mb > 5.27 $GeV/$c^2$, and
(c) $\mb$ distribution for $|\de| < 0.03$ GeV. Solid curves indicate
the fit results. }
\label{fg:alamc}
\end{figure}

Although the four- and three-body baryonic $B$ decays of
$\bz \to p\alamc \pi^+\pi^-$ and $\bp \to p\alamc \pi^+$ are experimentally
well observed\cite{Lamc}, there has no two-body mode being found.  
An effort is made to search for $\bz \to p\alamc$ with 
$\alamc \to \bar{p}K^+\pi^-$. Fig.~\ref{fg:alamc} shows the results. 
The measured branching 
fraction is ${\mathcal B}(B^0 \to p\alamc) = (2.19 ^{+0.56}_{-0.49} \pm 0.32
\pm 0.57) \times 10^{-5}$, where the first and the second errors are 
statistical and systematic, respectively. The last error is due to the 
uncertainty in the branching fraction $ {\mathcal B}(\alamc \to \bar{p}K^+\pi^-)
$\cite{PDG}. This is the first ever observation of a
two-body baryonic $B$ decays. One interesting feature is that this 
two-body decay rate is about factor of 10 smaller than that of 
the related three-body decay. In contrast, the two- and three-body mesonic $B$
decays are comparable.

\subsection{Results of $\bz \to \plpi$}

\begin{figure}[h]
\centering
\mbox{\epsfig{figure=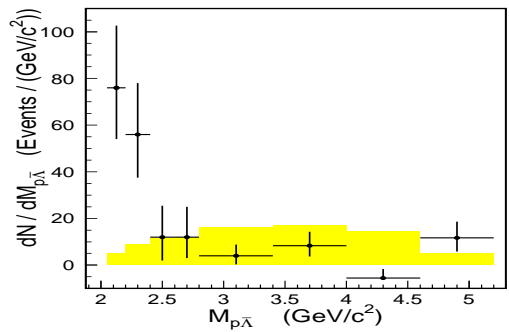,width=3in,height=2in}}
\centering
\caption{The fitted yield divided by the bin size for
$B^0\to \plpi$ as a function of $M_{p \bar{\Lambda}}$.
The shaded distribution is from a phase-space MC simulation 
with area normalized to
signal yield.}
\label{fg:phase}
\end{figure}

We observe a clear signal for  $\bz \to \plpi$, which is the
first hyperonic $B$ decay being found. 
Since the decay is not uniform in phase space,
we fit the $\Delta E$ signal yield in
bins of $M_{p \bar{\Lambda}}$,
and correct for the MC-determined detection
efficiency for each bin.
This reduces the  model dependence of the branching
fraction determination. 
The signal yield as a function
of $p\bar{\Lambda}$ mass is shown in Fig.~\ref{fg:phase}. 
This threshold peaking behavior is similar to that of 
$\bp \to \ppk$~\cite{ppk}.
The measured branching fraction is
${\mathcal B}(B^0 \to \plpi) = (3.97 ^{+1.00}_{-0.80} (stat.) \pm 0.56
(syst.) ) 
\times 10^{-6}$. 
Searches for $B^0 \to \plk$ and $\psigpi$ yield 
no significant signals 
and we set 90\% confidence-level upper limits of
${\mathcal B}(B^0 \to \plk) < 8.2 \times 10^{-7}$ and
${\mathcal B}(B^0 \to \psigpi) < 3.8 \times 10^{-6}$.

\subsection{Preliminary results of $B \to p\bar{p}h^{(*)}$}
Following our observation of $\bp\to p\bar{p}K^+$ mode, new
theoretical models tried to explain the 
experimental data and illustrate the possible 
correlations between similar decay modes. Some theoretical 
explanation~\cite{glue} about the 
threshold peaking behavior suggests possible glue-ball induced enhancement.
Table~\ref{pph} summarizes the search results of related three-body
decays. There is no evidence for possible glue-ball states in 2.2 GeV/$c^2 < 
\mpp <$ 2.4 GeV/$c^2$ region. 
These results are preliminary. Note that we apply charm veto~\cite{ppk} to 
these decay modes. However, the vetoed events with $\pp$ coming from $J/\psi$
decay can be used as a calibration tool. The determined branching fractions
using these events and the PDG~\cite{PDG} listed values are also shown in 
Table~\ref{pph} for comparison. 
 
\begin{table}[htb]
\caption{Summary of $p\bar{p}h^{(*)}$ results. Branching fraction products,   
${\mathcal B} (B \to J/\psi h^{(*)}) \times {\mathcal B} (J/\psi \to \pp)$,
are indicated with the $J/\psi$ symbol in the table. 
The search for
$\bp \to {\rm glue-ball} K^+, {\rm glue-ball} \to \pp$ yields an upper limit
at 90\% C.L. which is listed in the last row.    
}
\label{pph}
\vspace{0.4cm}
\begin{center}
\begin{tabular}{ccccc}
 &${\mathcal B}(\times 10^-6)$&Sig. &$J/\psi (\times 10^-6)$&
PDG $J/\psi(\times 10^-6$)
\\
\hline
$\ppk$&  $5.66^{+0.67}_{-0.57} \pm 0.62$&$15.3\sigma$&$2.48 \pm 0.25$&2.12
\\
$\pppi$& $3.06^{+0.73}_{-0.62} \pm 0.37$&$6.7\sigma$ &&
\\
$\ppks$& $0.94^{+0.53}_{-0.45} \pm 0.12$&$5.1\sigma$ &$1.25 \pm 0.25 $&0.92
\\
$\ppkst$&$10.31^{+3.52}_{-2.77} \pm 1.55$&$6.0\sigma$&$2.41 \pm 1.12$&2.94\\
$\ppksz$&$ < 7.6 \times 10^{-6}$ UL at 90\% C.L.&&$2.97 \pm 0.57$&2.78\\
glue-ball&$ < 0.9 \times 10^{-6}$ UL at 90\% C.L.
\\
\hline
\end{tabular}
\end{center}
\end{table}


\section*{Acknowledgments}
The author wish to thank the KEKB accelerator group for the excellent
operation of the KEKB accelerator and the Moriond QCD organization committee 
for making such a wonderful conference. This work is supported by the
National Science Council of the Republic of China under the grant
NSC-91-2112-M-002-028.

\end{document}